\documentclass[prl,aps,reprint,superscriptaddress,amsmath,amssymb]{revtex4-2}
\usepackage{graphicx,bm,mathptmx,physics,color,ulem}

\begin{document}

\title{Zeptosecond Angular Streak Camera}

\author{Hongcheng~Ni}
\email{hcni@lps.ecnu.edu.cn}
\affiliation{State Key Laboratory of Precision Spectroscopy, East China Normal University, Shanghai 200241, China}
\affiliation{Institute for Theoretical Physics, Vienna University of Technology, 1040 Vienna, Austria}
\affiliation{Collaborative Innovation Center of Extreme Optics, Shanxi University, Taiyuan, Shanxi 030006, China}
\affiliation{NYU-ECNU Joint Institute of Physics, New York University at Shanghai, Shanghai 200062, China}
\author{Stefan~Donsa}
\affiliation{Institute for Theoretical Physics, Vienna University of Technology, 1040 Vienna, Austria}
\author{Xiaochun~Gong}
\affiliation{State Key Laboratory of Precision Spectroscopy, East China Normal University, Shanghai 200241, China}
\affiliation{Collaborative Innovation Center of Extreme Optics, Shanxi University, Taiyuan, Shanxi 030006, China}
\author{Kiyoshi~Ueda}
\affiliation{State Key Laboratory of Precision Spectroscopy, East China Normal University, Shanghai 200241, China}
\affiliation{Department of Chemistry, Tohoku University, Sendai 980-8578, Japan}
\author{Jian~Wu}
\email{jwu@phy.ecnu.edu.cn}
\affiliation{State Key Laboratory of Precision Spectroscopy, East China Normal University, Shanghai 200241, China}
\affiliation{Collaborative Innovation Center of Extreme Optics, Shanxi University, Taiyuan, Shanxi 030006, China}
\affiliation{NYU-ECNU Joint Institute of Physics, New York University at Shanghai, Shanghai 200062, China}
\affiliation{CAS Center for Excellence in Ultra-intense Laser Science, Shanghai 201800, China}
\author{Joachim~Burgd\"orfer}
\email{joachim.burgdoerfer@tuwien.ac.at}
\affiliation{Institute for Theoretical Physics, Vienna University of Technology, 1040 Vienna, Austria}

\begin{abstract}
Time-resolved electronic processes on the attosecond scale have recently become experimentally accessible through the development of laser-based pump-probe interrogation techniques such as the attosecond streak camera, the reconstruction of attosecond beating by interference of two-photon transitions, and the attoclock. In this work, we demonstrate that by combining the concepts of the attosecond streak camera and the attoclock, time resolved processes down to the time scale of tens of zeptoseconds come into reach. Key to advancing to this remarkable level of time precision by this method termed the zeptosecond angular streak camera (ZASC) is its substantial intrinsic time-information redundancy. The ZASC results in a remarkably simple streaking trace, which is largely independent of the precise temporal structure of the streaking pulse, thereby bypassing the need for detailed characterization of the streaking field. Moreover, it is capable of retrieving information on the duration of the pump pulse. It is also capable of reaching attosecond-level precision in a single-shot mode that may be useful for free-electron-laser experiments. This concept promises to open pathways towards the chronoscopy of zeptosecond-level ultrafast processes.
\end{abstract}

\maketitle

The time resolution of electronic motion in atoms, molecules, and solids has become accessible through recent experimental advances. The production of ultrashort intense extreme ultraviolet (XUV) laser pulses \cite{paul01,hentschel01} from high harmonic generation (HHG) \cite{krause92,macklin93,popmintchev10} along with the development of spectroscopic techniques, such as the linear attosecond streak camera (LSC) \cite{hentschel01,itatani02}, the reconstruction of attosecond beating by interference of two-photon transitions (RABBITT) \cite{paul01}, and the attoclock technique \cite{eckle08a}, enables the temporal resolution in the few attosecond (1 as $=10^{-18}$ s) domain and, thus, on the natural time scale of electron dynamics. Recent time-resolved observations of the atomic photoionization \cite{schultze10,kluender11,su14}, tunneling ionization \cite{uiberacker07,eckle08b,pfeiffer12,torlina15,ni16,ni18a,ni18b,may21,sainadh19,kheifets20,hofmann21}, correlation-mediated photoionization time delay \cite{isinger17,ossiander17}, valence shell electron dynamics \cite{goulielmakis10,ma21}, and the laser-driven electron dynamics in a molecule \cite{odenweller11,cattaneo18,vos18,gong22} are striking examples for the capabilities of these techniques, leading to the rapid development of the attosecond metrology and chronoscopy \cite{pazourek15}.

The LSC uses a single isolated attosecond pulse mostly in the XUV spectral domain as the pump to photoionize an electron from the target and a short few-cycle laser pulse from infrared (IR) to terahertz (THz) as the probe to steer the energy of the emitted electron depending on its release time, thereby mapping time onto energy (classical energy streaking). The RABBITT technique typically employs longer attosecond pulse trains as the pump and a weaker multicycle IR pulse as the probe and achieves phase and temporal resolution by quantum path interference. Both the LSC and the RABBITT techniques employ a weak linearly polarized IR laser pulse as the probe and have been successfully used to time-resolve photoionization processes, specifically to determine time zero of photoionization. The attoclock, by contrast, employs a short circularly or elliptically polarized strong IR laser pulse acting both as the pump and the probe, thereby mapping the time of strong-field emission onto the deflection angle of the electron, referred to as angular streaking. Up to now, the attoclock approach has been primarily used for timing of strong-field tunneling processes.

Extensions of the attoclock protocol to a two-pulse pump-probe scheme have been realized by adding an XUV pulse as the pump and applied to characterize the carrier-envelope phase (CEP) of the XUV pulse \cite{he16,li17,fukahori17,garg18} and the temporal structure of a free-electron laser (FEL) pulse \cite{kazansky16,li18,hartmann18,kazansky19}. Very recently, such an angular streaking scheme has been applied to track the evolution of a coherent core-hole excitation in nitric oxide \cite{li22}.

In this Letter, we explore the merging of the concepts of the LSC and of the attoclock with the goal to realize precision timing, in particular of photoemission on an unprecedented (tens of) zeptosecond (1 zs $=10^{-21}$ s) scale. We refer to this merger as the zeptosecond angular streak camera (ZASC). Electronic processes on this novel ultrashort time scale have very recently come into focus \cite{grundmann20}: the time it takes light to travel from one atomic constituent to another within a molecule is about 250 zs. The photoemission from different constituents of the same molecule or from neighbouring atoms in a solid may thus acquire an additional structure-related time delay on the zeptosecond scale. Also, the Eisenbud-Wigner-Smith (EWS) time delay for photoionization of atoms can reach the zeptosecond scale for photon energies above a few hundred eV, e.g., for attosecond pulses from the FEL.

Differing from the conventional LSC, the ZASC employs a circularly or elliptically polarized streaking pulse where the angle, in addition to the energy, carries the timing information. Differing from the conventional attoclock, on the other hand, the ZASC employs an isolated sub-femtosecond pump pulse in the XUV or X-ray spectral regime, thus enabling the time-resolved study of photoionization of outer- or inner-shell electrons. By comparison with ab initio numerical solutions of the time-dependent Schr\"odinger equation (TDSE), we demonstrate that the ZASC can reach a time precision down to the zeptosecond level for the emission of moderately energetic photoelectrons ($E\gtrsim50$ eV). Key to achieving such a surprisingly high level of time precision is that this protocol takes advantage of the full vectorial momentum distribution. The built-in redundancy of information accessible for timing purpose supports such extremely high timing precision. In addition, we find that the ZASC is (within reasonable limits) largely insensitive to the detailed temporal structure of the streaking pulse, as well to the duration of the pump pulse, the latter can even be extracted from the streaking trace.

\begin{figure}
  \centering
  \includegraphics[width=\columnwidth]{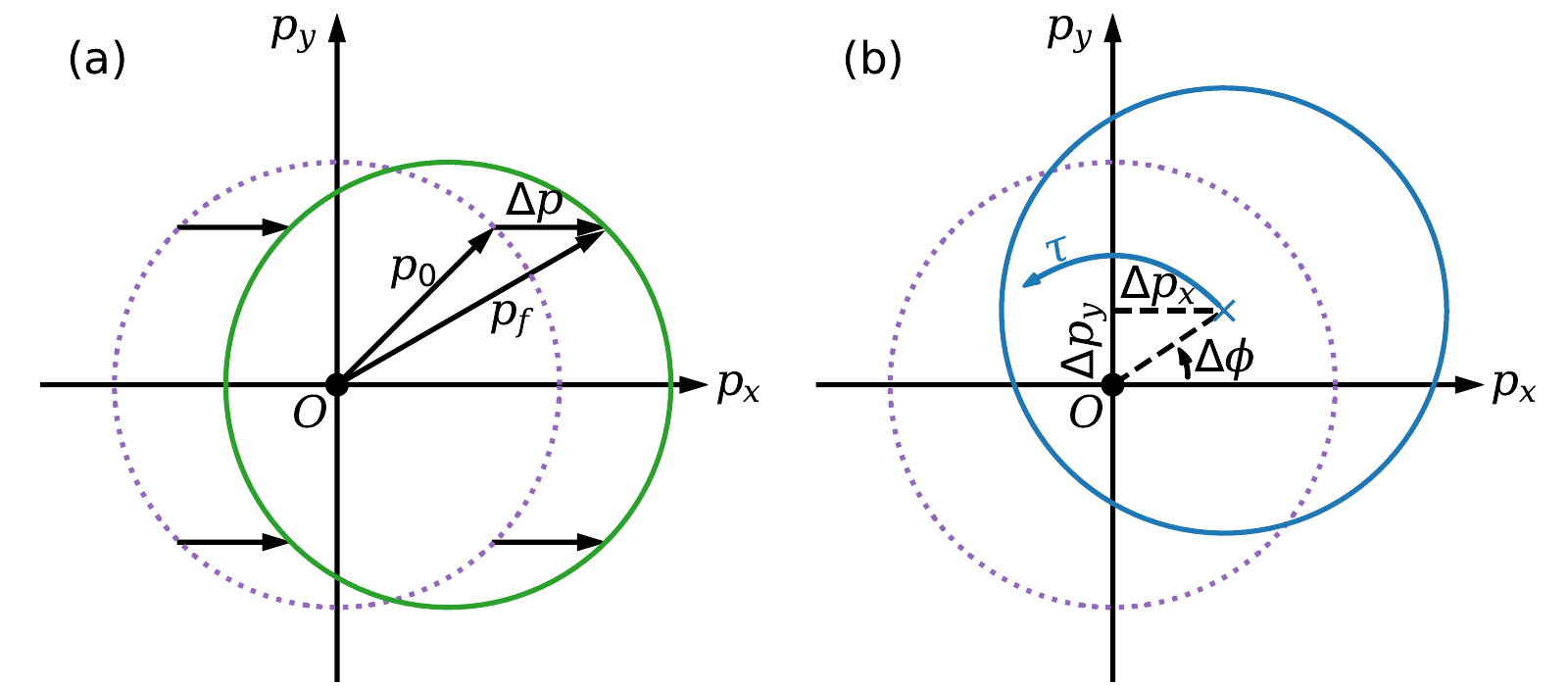}
  \caption{Comparison between the concepts of (a) the conventional linear attosecond streak camera (LSC) and (b) the zeptosecond angular streak camera (ZASC) schematically. The dashed circle represents the drift momentum of the photoelectron in the presence of the XUV pulse only. In panel (a), the LSC employs a linearly polarized IR streaking pulse which shifts the circle along the polarization direction with streaking momentum $\Delta p$. In panel (b), the ZASC makes use of a circularly (or elliptically) polarized streaking pulse, which rotates the center of the momentum distribution along a circle (or ellipse) when the delay $\tau$ between the attosecond pump and the probe pulse is varied.}
  \label{fig:scheme}
\end{figure}

We outline the concept of the ZASC by a comparison with the conventional LSC (Fig.~\ref{fig:scheme}). The simulations presented below employ a few-cycle streaking laser pulse with vector potential $\bm{A}(t)={A_0}\cos^4\left({\omega t}/{2N}\right)\left[-\cos(\omega t)\hat{\bm{e}}_x-\epsilon\sin(\omega t)\hat{\bm{e}}_y\right]$ polarized in the $x$--$y$ plane, where $A_0$ is the peak amplitude of the vector potential, $\epsilon$ is the pulse ellipticity, $\omega$ is the central frequency, and $N$ is the total number of cycles with $|\omega t|\leqslant N\pi$. The conventional LSC [Fig.~\ref{fig:scheme}(a)] employs a streaking pulse linearly polarized ($\epsilon=0$) in the $x$ direction. In the absence of the streaking pulse, the mean momentum of the photoelectron lies on a circle with radius $p_0=\sqrt{2E}$ (atomic units are used throughout unless noted otherwise), with $E$ the energy of the photoelectron, centered at the origin. When the streaking pulse is superimposed, the circle is displaced along the polarization direction by an amount of $\Delta p=-A(t_i)$ imparted by the vector potential $A$ of the streaking pulse at the ionization time $t_i$ of the photoelectron, thereby mapping photoemission time onto the streaked momentum $\Delta p$, and, hence, energy. In contrast, the present ZASC [Fig.~\ref{fig:scheme}(b)], employing a circularly or elliptically polarized streaking pulse, rotates the center of the circle in the the polarization plane, with the vectorial shifts $\Delta p_x=-A_x(t_i)$ and $\Delta p_y=-A_y(t_i)$ in the $x$ and $y$ directions, respectively. Thereby, the ionization time is mapped onto both the streaking angle
\begin{equation}
\Delta\phi=\arctan\left(\frac{\Delta p_y}{\Delta p_x}\right)=\arctan\left[\epsilon\tan(\omega t_i)\right]
\end{equation}
as well as on the magnitude of the mean momentum $\langle p\rangle$ (and energy). The streaking angle can be obtained from the photoelectron momentum distribution in the $x$--$y$ plane integrated over the $z$ direction.

To test the applicability and accuracy of the ZASC, we simulate a streaking experiment for photoionization of the hydrogen atom by numerically solving the TDSE. Photoionization of hydrogen is well studied with well established numerical and analytic solutions for the time zero of the ejection of the electronic wave packet \cite{pazourek15} against which our present computational streaking ``experiment'' can be benchmarked. The TDSE is solved using the well-established pseudospectral method \cite{tong97,tong17,gao19,ni20}. To compute the photoelectron momentum distribution, we project the wave function onto the Coulomb wave after the laser pulse is over \cite{supp}.

\begin{figure*}[t]
  \centering
  \includegraphics[width=\textwidth]{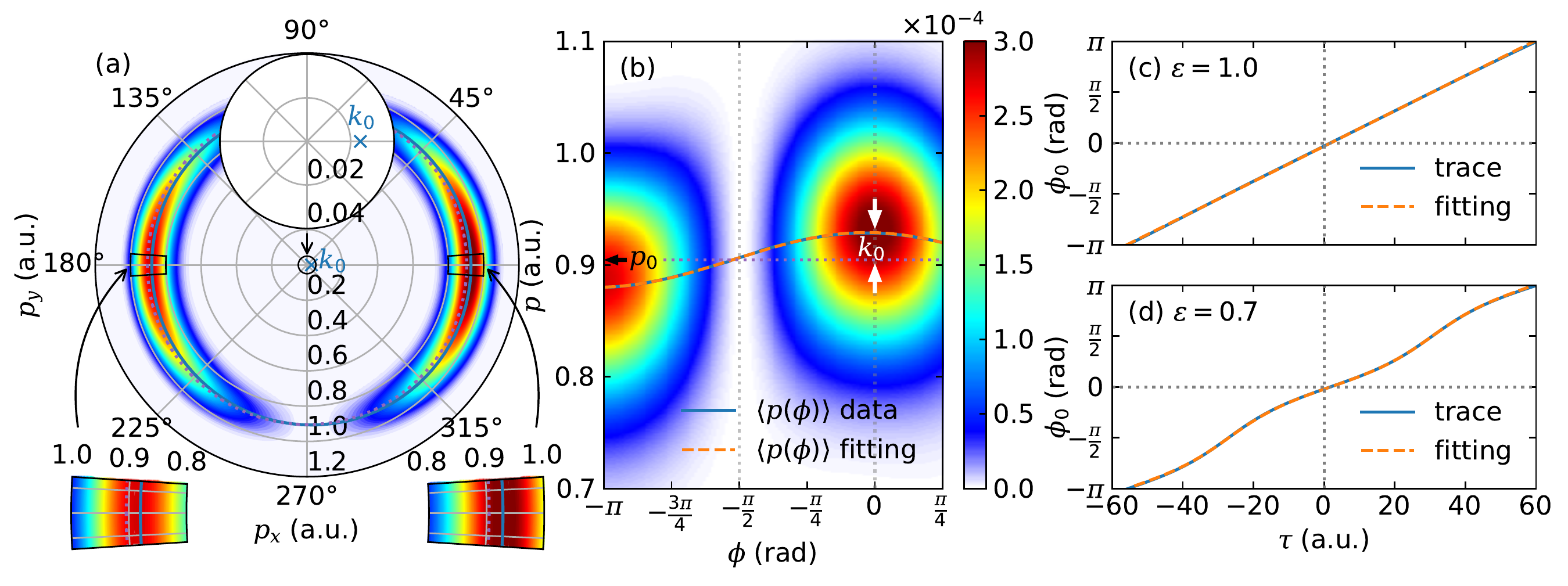}
  \caption{Photoelectron momentum distribution [panels (a) and (b)] with a circular streaking pulse. The peak of the photoelectron momentum distribution is located on a circle with radius $p_0$ (blue solid line) centered at $\bm{k}_0$, and is shifted relative to the IR-free case centered at origin (purple dotted line), see insets around $0^\circ$ and $180^\circ$. The center of the momentum distribution is obtained by fitting the angle-dependent expected radial momentum $\langle p(\phi)\rangle$ to Eq.~\eqref{eq:fitk0}. Streaking traces for circularly [panel (c), $\epsilon=1.0$] and elliptically [panel (d), $\epsilon=0.7$] polarized streaking pulses. In both cases, the streaking trace is monotonic and (near) linear. The corresponding streaking time delay $t_s$ is extracted from a fit to the trace [Eq.~\eqref{eq:fitts}].}
  \label{fig:trace}
\end{figure*}

To simulate the ZASC streaking trace, we scan the delay time $\tau$ between the XUV and IR pulses, $\bm{A}(t)=\bm{A}_\mathrm{IR}(t)+\bm{A}_\mathrm{XUV}(t-\tau)$. In the present work, we use a linearly polarized XUV pulse (polarized in the $x$ direction) with a peak intensity of $1\times10^{12}$ W/cm$^2$, a duration (full width at half maximum, FWHM) of $\tau_\mathrm{XUV}=522$ as (total duration of 2 fs under the cos$^4$ envelope), varying central frequencies $\omega_\mathrm{XUV}$, and a spectral width of $\Delta\omega_\mathrm{XUV}=0.15$ a.u. The IR pulse has a central wavelength of 800 nm corresponding to an optical period of $T_\mathrm{IR}=2.6$ fs, large compared to $\tau_\mathrm{XUV}$ ($T_\mathrm{IR}\gtrsim4\tau_\mathrm{XUV}$) as required for classical streaking \cite{pazourek15}. The total duration is two optical cycles ($N_\mathrm{IR}=2$). It is either circularly polarized ($\epsilon=1$) with a peak intensity of $2\times10^{11}$ W/cm$^2$ or elliptically polarized ($\epsilon=0.7$) with a peak intensity of $1.49\times10^{11}$ W/cm$^2$.

The projected two-dimensional photoelectron momentum distribution in the polarization plane (integrated over $p_z$) produced in photoionization of H by an XUV pump pulse with a central wavelength of 50 nm ($\omega_\mathrm{XUV}=0.91$) and a circular streaking pulse at an XUV-IR delay of $\tau=0$ [Fig.~\ref{fig:trace}(a)] displays a characteristic dipole $p$-wave pattern, its center however displaced from the origin by $\bm{k}_0\approx-\bm{A}_\mathrm{IR}(t=0)$ [Fig.~\ref{fig:trace}(a) inset]. Accordingly, the expectation value of the radial momentum $\langle p(\phi)\rangle$ varies with the azimuthal angle $\phi$ in the polarization plane and is shifted relative to its position in the absence of the streaking field. Its location is given by \cite{supp}
\begin{equation}
\langle p(\phi)\rangle=p_0+k_0\cos(\phi-\phi_0),
\label{eq:fitk0}
\end{equation}
where $p_0$ is the mean photoelectron momentum in the absence of the streaking field, $k_0$ is the magnitude of the IR-field-induced displacement of the photoelectron momentum distribution, and $\phi_0$ is the angle in the polarization plane, hereafter referred to as the streaking angle. The values of $p_0$, $k_0$, and $\phi_0$ can be obtained by fitting the momentum distribution to Eq.~\eqref{eq:fitk0} for the radial expectation value $\langle p(\phi)\rangle$. We emphasize that the full momentum distribution as input is crucial for achieving high accuracy in $\phi_0$ as it provides a high degree of redundancy of information.

\begin{figure*}[t]
  \centering
  \includegraphics[width=\textwidth]{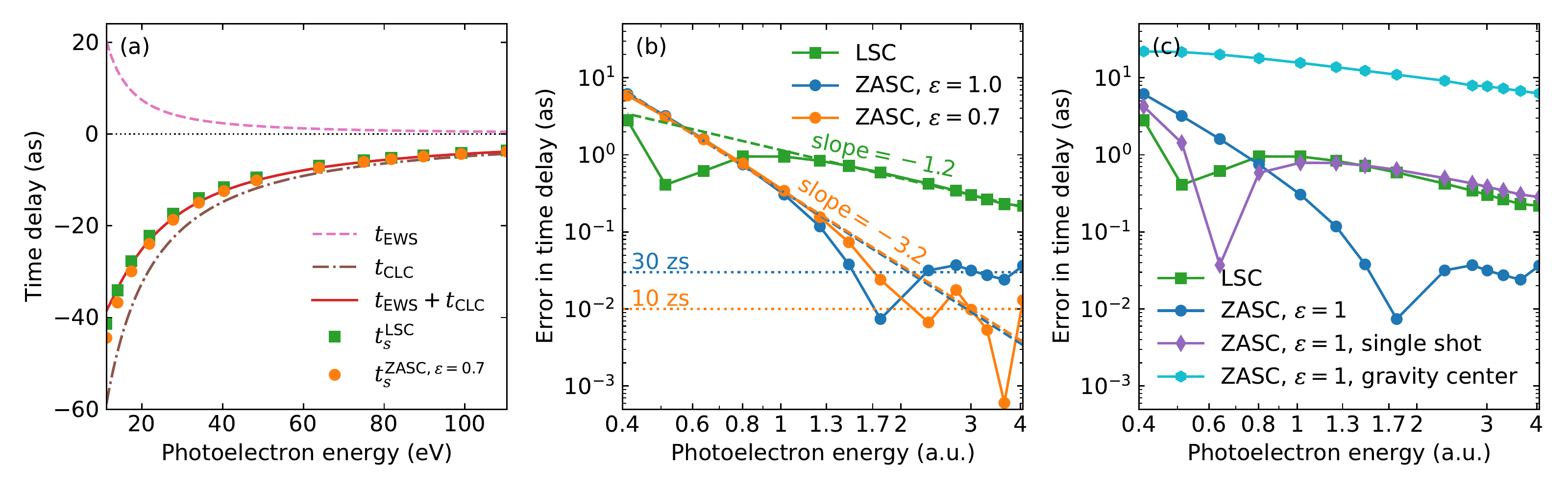}
  \caption{(a) Photoionization time delays of the hydrogen atom. The pink dashed line is the EWS time delay $t_\mathrm{EWS}$, the brown dash-dotted line is the CLC delay $t_\mathrm{CLC}$, and their sum is the streaking time delay (red solid line). The numerical data points are from the simulations of the conventional LSC $t_s^\mathrm{LSC}$ (green squares) and the ZASC $t_s^{\mathrm{ZASC},\epsilon=0.7}$ (orange dots). (b) Estimates of the error in the streaking delay using different streaking protocols. The conventional LSC (green line with squares) has an error on the 1 as scale for low $E$ ($E\sim1$ a.u.), and is reduced for higher $E$ as $\sim E^{-1.2}$. By contrast, the error of the ZASC rapidly diminishes as $\sim E^{-3.2}$ and reaches $\lesssim30$ zs already at $E\approx1.8$ a.u.\ (or 50 eV) for both elliptically polarized ($\epsilon=0.7$, orange line) and circularly ($\epsilon=1.0$, blue line) polarized streaking field. (c) Error in the streaking delay when the information redundancy is partially removed by either using a single-shot scenario (purple line with diamonds) or using the center of gravity of the photoelectron momentum distribution (cyan line with hexagons) without taking advantage of the information of the full photoelectron momentum distribution.}
  \label{fig:delay}
\end{figure*}

Scanning now the XUV-IR time delay $\tau$ thereby generating a streaking trace, the streaking angle varies as
\begin{equation}
\phi_0(\tau)=\arctan\left[\epsilon\tan\left(\omega_s(\tau+t_s)\right)\right].
\label{eq:fitts}
\end{equation}
This remarkably simple relation is fulfilled with high accuracy for both elliptic polarization [Fig.~\ref{fig:trace}(d)] as well as circular polarization [Fig.~\ref{fig:trace}(c)]. In the latter case, the streaking trace further reduces to a straight line $\phi_0(\tau)=\omega_s(\tau+t_s)$. The streaking time shift $t_s$ and, hence, the EWS time delay of photoionization \cite{pazourek15} becomes now directly accessible through a fit to Eq.~\eqref{eq:fitts}. One important point to note is that the effective frequency $\omega_s$ with which the streaking angle rotates does not, in general, coincides with that of the streaking field $\omega_\mathrm{IR}$. Instead, the streaking trace entails a convolution of the ionization process with the IR angular streaking resulting in a modified frequency approximately given by \cite{supp}
\begin{equation}
\omega_s=\frac{1}{1+\left(\tau_\mathrm{XUV}/\tau_\mathrm{IR}\right)^2/2}\omega_\mathrm{IR}.
\label{eq:w}
\end{equation}
The fit to the recorded streaking trace, therefore, allows, in addition, to determine the ratio of XUV and IR pulse duration $\tau_\mathrm{XUV}/\tau_\mathrm{IR}$. The latter may provide an alternative to protocols for determining FEL pulse durations \cite{wieland21} and aid in their characterization.

In general, the streaking time delay can be written as a sum of the Eisenbud-Wigner-Smith (EWS) time delay $t_\mathrm{EWS}$ \cite{eisenbud48,wigner55,smith60} and the Coulomb-laser-coupling (CLC) delay $t_\mathrm{CLC}$ \cite{pazourek15}. For the hydrogen atom, the EWS time delay is given analytically by $t_\mathrm{EWS}(Z,E,l) = \frac{Z}{(2E)^{3/2}}\Re\left[\Psi\left(1+l-i\frac{Z}{\sqrt{2E}}\right)\right]$, where $Z=1$ is the asymptotic charge of the residual ion, $E$ is the photoelectron energy, $l=1$ is the angular quantum number of the photoelectron, and $\Psi$ is the digamma function, while the CLC delay is written as $t_\mathrm{CLC}(Z,E,\omega_\mathrm{IR}) = \frac{Z}{(2E)^{3/2}}\left[2-\ln\left(2\pi E/\omega_\mathrm{IR}\right)\right]$ \cite{pazourek15}. The corresponding quantity for the RABBITT protocol is usually referred to as the continuum-continuum delay \cite{dahlstroem12,dahlstroem13} and can be identified as the EWS delay for continuum-continuum transitions \cite{fuchs20}. For photon energies well above the ionization threshold ($E\gtrsim1$ a.u.), the sum of $t_\mathrm{EWS}$ and $t_\mathrm{CLC}$ predicts the streaking delay with high accuracy, against which our numerical simulation of the streaking ``experiment'' employing either the conventional LSC or the ZASC will be benchmarked.

The theoretically predicted streaking time delay $t_s=t_\mathrm{EWS}+t_\mathrm{CLC}$ is, on the attosecond time scale, quite well reproduced over a wide range of photon energies by the streaking simulation for the LSC $t_s^\mathrm{LSC}$ and the ZASC $t_s^\mathrm{ZASC}$ [Fig.~\ref{fig:delay}(a)]. For the latter, we have used an elliptical streaking pulse with $\epsilon=0.7$. For assessing now the precision of these streaking protocols, we evaluate the error $\mathcal{E}$ given by the deviation of the streaking time from the theoretical prediction [Fig.~\ref{fig:delay}(b)]. With increasing photoelectron energy, the error of the conventional LSC $\mathcal{E}^\mathrm{LSC}$ is reduced, with a slope of about $-1.2$, i.e., $\mathcal{E}^\mathrm{LSC}\sim E^{-1.2}$, while the error of the ZASC drops precipitously with a slope of around $-3.2$, i.e., $\mathcal{E}^\mathrm{ZASC}\sim E^{-3.2}$, reaching a level of 30 zs for the circular ZASC and 10 zs for the elliptical ZASC at $E\approx1.8$ (or 50 eV). Clearly, for sufficiently high photoelectron energies, the ZASC is much more precise than the conventional LSC, and promises to clock ultrafast dynamical processes down to the zeptosecond level.

To unravel the origin of this remarkable level of precision, a closer comparison with the conventional LSC can provide key insights and useful guidance. Already the attosecond time resolution of the LSC defies the notion that the time resolution of a camera is limited by the ``duration of the shutter''. The duration of both the pump pulse (typically $\sim500$ as for isolated XUV pulses) and of the probe pulse with a single cycle period $\gtrsim2.5$ fs, exceeds the achieved single-digit attosecond resolution by (at least) two orders of magnitude. (Analogous relations apply to the RABBITT protocol.) Nevertheless, the attosecond precision emerges from the fact that the energy of the streaked electron varies sinusoidally with the linearly polarized vector potential whose oscillation time dependence $\sim\cos(\omega_st+\phi_0)$ is well known and well characterized. Accurate determination of the phase angle $\phi_0$ from a fit to the streaking trace with an error in the phase angle of the order of $\Delta\phi_0\lesssim\pi/100$ is straightforward and results in a time resolution by $\sim2$ orders of magnitude shorter than the period of the streaking field. The present ZASC extends now this extraction of accurate phase information via the known vector potential from the scalar $E$ to the two-dimensional vectorial momentum $\langle\bm{p}(\phi)\rangle$ in the polarization plane, by independent fits to $\cos(\omega_st+\phi_0)$ and $\sin(\omega_st+\phi_0)$ when the full vectorial momentum distribution is available. This increased redundancy of information results in an increase of precision by another two orders of magnitude as illustrated in Fig.~\ref{fig:delay}(b). The XUV-IR delay scan of the full vectorial momentum distribution invoked by the present ZASC enables to effectively suppress the error, promising a zeptosecond chronoscopy with presently available laser settings.

The high information redundancy acquired by the ZASC has the consequence that even when the experimentally accessible information is significantly reduced, remarkable time resolution (even though not on the zeptosecond scale) is still within reach. We illustrate this point with two examples. First, the ZASC can even operate in a single-shot mode where a scan of the XUV-IR delay is omitted. For a given XUV-IR delay $\tau$, the streaking time delay $t_s$ can be obtained directly from Eq.~\eqref{eq:fitts} when the (small) deviation of $\omega_s$ from $\omega_\mathrm{IR}$ is neglected. In this limit, $t_s\approx\arctan[\tan\phi_0(\tau)/\epsilon]/\omega_\mathrm{IR}-\tau$. The error of such a single-shot protocol is obtained as an average over the different $\tau$ values within an optical cycle of the streaking pulse as $\tau$ may be ill defined due to the temporal jitter of the XUV pump pulse. Even in this single-shot mode the ZASC features a precision competitive with the conventional LSC [Fig.~\ref{fig:delay}(c)]. Second, rather than taking advantage of the full photoelectron momentum distribution when fitting the streaking angle $\phi_0$ [Eq.~\eqref{eq:fitk0}] for a given XUV-IR delay $\tau$, one can employ only the center of gravity of the momentum distribution $\langle\bm{p}\rangle$ which is a highly integrated observable that contains a much reduced amount of information. Nevertheless, the resulting streaking trace generated by scanning the XUV-IR delay $\tau$ yields the information with the precision on the $\sim10$ as scale [Fig.~\ref{fig:delay}(c)].

Furthermore, the simplicity of the extracted trace [Eq.~\eqref{eq:fitts}] is (within reasonable limits) independent of the temporal structure of the streaking laser pulse, in contrast to the conventional LSC. This fact results in a high flexibility of the ZASC, making it applicable without detailed a priori knowledge of the streaking pulse and may help to avoid complications with the characterization of the streaking pulse. By varying the envelope of the streaking pulse we have verified that its influence on the time resolution is small \cite{fuchs20}.

To experimentally carry out the ZASC protocol, an attosecond coincidence streaking camera \cite{sabbar14} featuring a two-dimensional photoelectron momentum spectrometer in conjunction with a table-top source of XUV isolated attosecond pulses and phase controlled pump-probe optical beamline is required. Alternatively, the coaxial velocity map imaging technique \cite{li22} that has been applied in FEL settings could be used. Both could pave the way to realize the zeptosecond resolution in time-resolved quantum dynamics.

In summary, we have proposed the concept of a zeptosecond angular streak camera (ZASC) which employs an XUV pump pulse and a circularly or elliptically polarized IR streaking pulse for precision timing of the photoionization process. Unlike the conventional linear attosecond streak camera (LSC), it promises a time precision down to the level of a few tens of zeptoseconds. This unprecedented resolution results from the high information redundancy encoded in the streaked vectorial momentum distribution, improving upon the standard LSC by another two orders of magnitude. The present ZASC also has the potential to realize attosecond time resolution when only significantly reduced information is accessible, in particular it may operate in a single-shot mode which could be helpful for future FEL experiments where time jitter is unavoidable. It also allows to extract the XUV pulse duration. The time resolution achievable by the ZASC offers access to new classes of ultrafast processes on the zeptosecond scale. They include inner-shell atomic processes, photon-propagation-related time delays between different atomic constituents of molecules and solids, and observation in inner-shell transitions involving the coupling of electronic and nuclear degrees of freedom such as nuclear electron capture.

This work was supported by the National Key R\&D Program of China (Grant No.~2018YFA0306303), the National Natural Science Foundation of China (Grant Nos.~92150105, 11904103, 11834004), the Austrian Science Fund (Grant Nos.~M2692, W1243), and the Science and Technology Commission of Shanghai Municipality (Grant Nos.~21ZR1420100, 19JC1412200). Numerical computations were performed on the ECNU Multifunctional Platform for Innovation (001) and the Vienna Scientific Cluster (VSC).

\end{document}